\documentclass[prb,aps,twocolumn,amsmath,amssymb,floatfix,showpacs,
superscriptaddress]{revtex4}
\usepackage[dvips]{graphics}

\usepackage{color}
\definecolor{dred}{rgb}{0.4,0,0.2}

\begin{document}

\title{Positional dependence of energy gap on line defect in armchair 
graphene nanoribbons: Two-terminal transport and related issues}

\author{Paramita Dutta}

\affiliation{Theoretical Condensed Matter Physics Division, Saha 
Institute of Nuclear Physics, Sector-I, Block-AF, Bidhannagar, 
Kolkata-700 064, India}

\author{Santanu K. Maiti}

\email{santanu.maiti@isical.ac.in}

\affiliation{Physics and Applied Mathematics Unit, Indian Statistical
Institute, 203 Barrackpore Trunk Road, Kolkata-700 108, India}

\author{S. N. Karmakar}

\affiliation{Theoretical Condensed Matter Physics Division, Saha
Institute of Nuclear Physics, Sector-I, Block-AF, Bidhannagar,
Kolkata-700 064, India}

\begin{abstract}
The characteristics of energy band spectrum of armchair graphene 
nanoribbons in presence of line defect are analyzed within a simple 
non-interacting tight-binding framework. In metallic nanoribbons an
energy gap may or may not appear in the band spectrum depending on the
location of the defect line, while in semiconducting ribbons the gaps 
are customized, yielding the potential applicabilities of graphene
nanoribbons in nanoscale electronic devices. With a more general model, 
we also investigate two-terminal electron transport using Green's function 
formalism.
\end{abstract}


\maketitle

\section{Introduction}

The fabrication of single layer graphene by Novoselov {\it et 
al.}~\cite{novoselov} in 2004 has opened a new era in the research of 
low-dimensional nanostructured materials. Graphene, the carbon allotrope 
with planar honeycomb lattice structure, is a promising candidate of 
nano-electronic components owing to its exceptional electronic, thermal 
and transport properties~\cite{neto}. It has been predicted that graphene 
sheet is a zero gap semiconductor~\cite{abergel}, but its behavior strongly 
depends on the boundary conditions when it is tailored into ribbon, flake 
or tube~\cite{we1}. The sensitivity to the ribbon width, chirality, shape 
of the edges has allowed one to switch its semiconductor-like behavior 
from zero gaped to the finite one. Intensive researches have already been 
done on graphene nanoribbons (GNRs) to explore the influence of edge 
topology~\cite{barone,nakada} on transport properties. Some attempts 
including chemical doping~\cite{ouyang,huang}, application of uniaxial 
strain~\cite{sena,chang}, chemical edge modifications~\cite{wang}, 
incorporation of impurity~\cite{tsuyuki}, line defect~\cite{lahiri,lin,
okada,gunlycke,costa} are in focus of study on this system. But to realize 
the potential application of this material the control over transport 
properties needs to be clarified in a deeper way.
 
To date, many theoretical~\cite{fujita, waka,schulz} as well as 
experimental~\cite{han,cooper} works have been done which reveal the fact 
that graphene nanoribbons (GNRs) with zigzag edges exhibit a metallic phase
with localized states located on the edges, while armchair graphene 
nanoribbons (AGNRs) show metallic or semiconducting phase depending on 
the width of the ribbons~\cite{zheng,son,chen}. This phenomenon is true only 
for clean ribbons i.e., without any deformation anywhere in the sample. 
But, the presence of impurity or deformation makes the system behave 
differently. In 2007 Peeters {\it et al.}~\cite{costa} have shown that 
in presence of line impurity in graphene nanoribbon a gap opens up in the 
energy band spectrum. The system they considered was practically coupled 
two graphene ribbons of different sizes separated by a distance. Line defect 
also yields the possibility of using graphene as a valley filter as 
demonstrated by Gunlycke {\it et al.}~\cite{gunlycke}. In a recent experiment 
topological line defect has been studied using scanning tunneling 
microscope~\cite{lahiri,okada}. Although the studies involving AGNRs have
already generated a wealth of literature there is still need to look deeper
into the problem to address several important issues those have not been
\begin{figure}[ht]
{\centering \resizebox*{8.2cm}{4.5cm}{\includegraphics{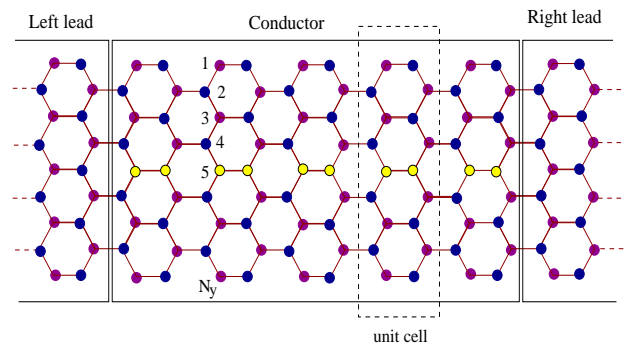}}\par}
\caption{(Color online). A schematic illustration of an armchair graphene 
nanoribbon, coupled to left and right leads, in presence of line defect 
(yellow circles). The blue and magenta circles represent two different 
sub-lattices of the ribbon.}
\label{model}
\end{figure}
well explored earlier, as for examples the understanding of the behavior of
energy gap in metallic and semiconducting AGNRs in presence of line defect
and also the dependence of this energy gap on the location of line defect. 
In the present work we mainly concentrate on these issues. Here we analyze 
the energy band structure of simple armchair graphene nanoribbons in presence 
of line defect within a tight-binding model. We show that depending on the
location of a line defect in a metallic AGNR an energy gap may or may not
appear in the band spectrum, while in a semiconducting AGNR the gap can be 
controlled. This phenomenon leads to the possibility of using AGNRs as 
nanoscale electronic devices. With a more specific model we also describe 
two-terminal electron transport through finite sized AGNRs by attaching
them to two semi-infinite graphene leads using Green's function formalism 
to explore the results in realistic cases.

The structure of the paper is as follows. In Section II, we describe the 
specific model i.e., AGNR with side-attached leads and theoretical 
formulation to illustrate two-terminal electronic transport. The essential
results are presented in Section III which contains (a) energy band 
structure of simple isolated armchair graphene nanoribbons in presence 
of line defect, and (b) transmission probability as a function of injecting 
electron energy through the lead-AGNR-lead bridge system. Finally, in 
Section IV, we summarize our main results and discuss their possible 
applications for further study.

\section{Model and theoretical formulation}

{\bf \underline{Model}:} Figure~\ref{model} presents a schematic illustration
of the model quantum system, where a finite size AGNR is coupled to two
semi-infinite graphene leads with armchair edges. The blue and magenta 
circles correspond to two different sublattices of the ribbon, and, the
yellow circles, representing the defect sites, are arranged in a line
result a defect line. A unit cell of the AGNR is described by the dashed
region which contains $2 N_y$ atomic sites in our notation. 
 
Our analysis for the present work is based on non-interacting electron 
picture, and, within this framework, tight-binding (TB) model is extremely
suitable for analyzing electron transport through such a two-terminal
bridge system. The single particle Hamiltonian which captures the AGNR
and side-attached leads gets the form:
\begin{equation}
\mbox{\boldmath$H$}=\mbox{\boldmath$H$}_{\mbox{\tiny AGNR}} + 
\mbox{\boldmath$H$}_{\mbox{\tiny lead}} + 
\mbox{\boldmath$H$}_{\mbox{\tiny tun}}. 
\label{eq1}
\end{equation}
The first term $\mbox{\boldmath$H$}_{\mbox{\tiny AGNR}}$ denotes the 
Hamiltonian of the AGNR sandwiched between two graphene leads. Under
nearest-neighbor hopping approximation, the TB Hamiltonian of the AGNR
reads,
\begin{equation}
\mbox{\boldmath$H$}_{\mbox{\tiny AGNR}}= \sum_l \epsilon_l c_l^{\dag} c_l +
\sum_l v_{l,l+1} \left(\mbox{\boldmath$c$}^{\dag}_l \mbox{\boldmath$c$}_{l+1} 
+ \mbox{h.c.} \right)
\label{hr}
\end{equation}
where, $\epsilon_l$ is the on-site energy and $v_{l,l+1}$ is the 
nearest-neighbor hopping integral. The hopping integral $v_{l,l+1}$ is 
set equal to $v$ or $v^{\prime}$ whether an electron hops between two ordered
atomic sites or between two defect sites, and it is $v^{\prime \prime}$
when the hopping of an electron takes place between an ordered and a defect
site. $c_l^{\dagger}$ ($c_l$) is the creation (annihilation) operator of 
an electron at the site $l$.

The second and third terms of Eq.~\ref{eq1} correspond to the Hamiltonians
for the semi-infinite graphene leads (left and right leads) and
AGNR-to-lead coupling. A similar kind of TB Hamiltonian (see Eq.~\ref{hr})
is used to illustrate the leads where the Hamiltonian is parametrized by 
constant on-site potential $\epsilon_0$ and nearest-neighbor hopping
integral $v_0$. The AGNR is directly coupled to the leads by the parameters
$\tau_L$ and $\tau_R$, where they (coupling parameters) correspond to the
coupling strengths between the edge sites of the ribbon and the left and 
right leads, respectively.

{\bf \underline{Transmission probability}:} To obtain 
electronic transmission probability through the AGNR we use Green's 
function formalism. Within the regime of coherent transport and in the
absence of Coulomb interaction this technique is well applied. Using
Fisher-Lee relation, two-terminal transmission probability $T$ through
the lead-AGNR-lead bridge system can be written as~\cite{datta1},
\begin{eqnarray}
T = {\mbox{Tr}} \left[\mbox{\boldmath$\Gamma$}_L  
\mbox{\boldmath$G$}^r_{\mbox{\tiny AGNR}} \mbox{\boldmath$\Gamma$}_R  
\mbox{\boldmath$G$}^a_{\mbox{\tiny AGNR}} \right]
\label{transmission}
\end{eqnarray}
where, $\mbox{\boldmath$\Gamma$}_L$ and $\mbox{\boldmath$\Gamma$}_R$ 
are the coupling matrices and $\mbox{\boldmath$G$}^r_{\mbox{\tiny AGNR}}$ 
and $\mbox{\boldmath$G$}^a_{\mbox{\tiny AGNR}}$ are the retarded and 
advanced Green's functions of the AGNR, respectively. Now the single
particle Green's function operator representing the entire system for 
an electron with energy $E$ is defined as,
\begin{eqnarray} 
\mbox{\boldmath$G$}=\left[(E+i\eta)\mbox{\boldmath$I$}-\mbox{\boldmath$H$} 
\right]^{-1}
\label{green}
\end{eqnarray}
where, $\eta \rightarrow 0^+$, $\mbox{\boldmath$H$}$ represents the 
Hamiltonian of the full system and $\mbox{\boldmath$I$}$ is the identity
matrix. Following the matrix forms of $\mbox{\boldmath$H$}$ and 
$\mbox{\boldmath$G$}$, the problem of finding $\mbox{\boldmath$G$}$ in
the full Hilbert space of $\mbox{\boldmath$H$}$ can be mapped exactly
to a Green's function $\mbox{\boldmath$G$}_{\mbox{\tiny AGNR}}$ 
corresponding to an effective Hamiltonian in the reduced Hilbert space
of the AGNR itself and we get~\cite{datta1},
\begin{eqnarray}
\mbox{\boldmath$G$}_{\mbox{\tiny AGNR}}=\left[(E+i\eta)\mbox{\boldmath$I$} 
-\mbox{\boldmath$H$}_{\mbox{\tiny AGNR}}-\mbox{\boldmath$\Sigma$}_L 
- \mbox{\boldmath$\Sigma$}_R \right]^{-1}.
\label{greendevice}
\end{eqnarray}
Here, $\mbox{\boldmath$\Sigma$}_L$ and $\mbox{\boldmath$\Sigma$}_R$ 
are the contact self-energies introduced to incorporate the effect of
coupling of the AGNR to the left and right leads, respectively. Below
we represent explicitly how these self-energies are evaluated for the
graphene leads attached to the AGNR.

{\bf{\underline {Evaluation of self-energy}}:} In order to determine 
self-energies for these side-attached leads we follow the prescription
\begin{figure}[ht]
{\centering \resizebox*{8.6cm}{1.6cm}{\includegraphics{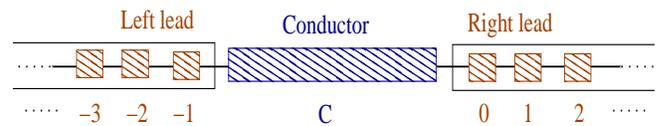}}\par}
\caption{(Color online). A schematic view of our model quantum system where
AGNR and side-attached leads are sketched with discrete principal layers 
those are specified by integer numbers.}
\label{unit}
\end{figure}
addressed by Sancho {\it et al.}~\cite{sancho}, where both leads and
AGNR are sketched with discrete effective principal layers. These layers 
are defined as the smallest group of neighboring atomic planes and they 
allow only nearest-neighbor interaction between them. It effectively 
transforms the original system into a linear chain of principal 
layers~\cite{nardelli} as shown in Fig.~\ref{unit}. We label the principal 
layers in the right-lead as $0$, $1$, $2$, $\dots$ and in the left-lead as 
$-1$, $-2$, $-3$, $\dots$ and so on. The sample between the leads is 
denoted by $C$. Below, we describe elaborately the evaluation of the 
self-energy corresponding to the right lead, and, following this 
prescription we also determine the self-energy for the left lead. Using 
Eq.~\ref{green} a set of equations for the layer orbitals can be written as,
\begin{eqnarray}
\left[(E+i \eta) \mbox{\boldmath$I$}- \mbox{\boldmath$H$}_{0,0}\right]
\mbox{\boldmath$G$}_{0,0}& =& \mbox{\boldmath$I$} +  
\mbox{\boldmath$H$}_{0,1} \mbox{\boldmath$G$}_{1,0} \nonumber \\
\left[(E+i \eta) \mbox{\boldmath$I$}- \mbox{\boldmath$H$}_{0,0}\right]
\mbox{\boldmath$G$}_{1,0}& =& \mbox{\boldmath$H$}_{0,1}^{\dag} 
\mbox{\boldmath$G$}_{0,0} +  \mbox{\boldmath$H$}_{0,1} 
\mbox{\boldmath$G$}_{2,0} \nonumber \\   
\left[(E+i \eta) 
\mbox{\boldmath$I$}- \mbox{\boldmath$H$}_{0,0}\right] 
\mbox{\boldmath$G$}_{2,0}&=&  \mbox{\boldmath$H$}_{0,1}^{\dag}  
\mbox{\boldmath$G$}_{1,0}+  \mbox{\boldmath$H$}_{0,1}  
\mbox{\boldmath$G$}_{3,0} \nonumber \\ 
~~~~~~~~~~~~\hdots \nonumber \\
\left[(E+i \eta) \mbox{\boldmath$I$}- \mbox{\boldmath$H$}_{0,0} \right] 
\mbox{\boldmath$G$}_{n,0}&=&  \mbox{\boldmath$H$}_{0,1}^{\dag} 
 \mbox{\boldmath$G$}_{n-1,0}+  \mbox{\boldmath$H$}_{0,1}
 \mbox{\boldmath$G$}_{n+1,0} \nonumber \\  
\label{greeneq}
\end{eqnarray}
where, we assume that 
$\mbox{\boldmath$H$}_{0,0}=\mbox{\boldmath$H$}_{1,1}=
\mbox{\boldmath$H$}_{2,2}=\dots$ and $\mbox{\boldmath$H$}_{0,1}=
\mbox{\boldmath$H$}_{1,2}=\mbox{\boldmath$H$}_{2,3}=\dots$. Here, 
$\mbox{\boldmath$H$}_{l,l}$ describes the Hamiltonian of $l$-th principal 
layer, while $\mbox{\boldmath$H$}_{l,l+1}$ corresponds to the coupling 
matrix between $l$-th and ($l+1$)-th layers. The general expression of 
$\mbox{\boldmath$G$}_{n,0}$ reads,
\begin{eqnarray}
\mbox{\boldmath$G$}_{n,0}&=&\mbox{\boldmath$t$}_0\mbox{\boldmath$G$}_{n-1,0}
+ \mbox{\boldmath$\tilde{t}$}_0 \mbox{\boldmath$G$}_{n+1,0} 
\label{gn0}
\end{eqnarray}
where, 
\begin{eqnarray}
\mbox{\boldmath$t$}_0&=&[(E+i \eta)\mbox{\boldmath$I$} 
-\mbox{\boldmath$H$}_{0,0}]^{-1} \mbox{\boldmath$H$}_{0,1}^{\dag},\nonumber\\
\mbox{\boldmath$\tilde{t}$}_0 & = &[(E+i \eta) \mbox{\boldmath$I$} 
-\mbox{\boldmath$H$}_{0,0}]^{-1} \mbox{\boldmath$H$}_{0,1}.
\end{eqnarray}
Substituting $\mbox{\boldmath$G$}_{n,0}$ into the expressions of 
$\mbox{\boldmath$G$}_{n-1,0}$ and $\mbox{\boldmath$G$}_{n+1,0}$ 
we can write, 
\begin{eqnarray}
\mbox{\boldmath$G$}_{n,0}&=& \mbox{\boldmath$t$}_1 \mbox{\boldmath$G$}_{n-2,0}
+ \mbox{\boldmath$\tilde{t}$}_1 \mbox{\boldmath$G$}_{n+2,0} ~~~~~ (n \ge 2)
\label{gn}
\end{eqnarray}
and this process continues iteratively to repeat. After $i$-th iteration 
we have,
\begin{eqnarray}
\mbox{\boldmath$G$}_{n,0}&=& \mbox{\boldmath$t$}_i 
\mbox{\boldmath$G$}_{n-2^i,0}+ \mbox{\boldmath$\tilde{t}$}_i 
\mbox{\boldmath$G$}_{n+2^i,0} ~~~~~ (n \ge 2^i)
\label{gni}
\end{eqnarray}
with,
\begin{eqnarray}
\mbox{\boldmath$t$}_{i}=(\mbox{\boldmath$I$} -\mbox{\boldmath$t$}_{i-1}
\mbox{\boldmath$\tilde{t}$}_{i-1}-\mbox{\boldmath$\tilde{t}$}_{i-1} 
\mbox{\boldmath$t$}_{i-1})^{-1} \mbox{\boldmath$t$}_{i-1}^2, 
\nonumber \\
\mbox{\boldmath$\tilde{t}$}_{i}=(\mbox{\boldmath$I$} -
\mbox{\boldmath$t$}_{i-1} \mbox{\boldmath$\tilde{t}$}_{i-1}-
\mbox{\boldmath$\tilde{t}$}_{i-1} \mbox{\boldmath$t$}_{i-1})^{-1} 
\mbox{\boldmath$\tilde{t}$}_{i-1}^2.
\end{eqnarray}
The iteration is to be done until $\mbox{\boldmath$t$}_{i}$, 
$\mbox{\boldmath$\tilde{t}$}_i \le \delta$ with $\mbox{\boldmath$\delta$}$ 
arbitrarily small.

Using these $\mbox{\boldmath$t$}_i$'s we can determine the Green's 
function of a single layer in terms of the Green's function of the 
following or preceding one like,
\begin{eqnarray}
\mbox{\boldmath$G$}_{1,0}=\mbox{\boldmath$T$} \mbox{\boldmath$G$}_{0,0} 
~~~~~\mbox{and} ~~~~~~~~
\mbox{\boldmath$G$}_{0,0}=\mbox{\boldmath$\tilde{T}$} 
\mbox{\boldmath$G$}_{1,0}
\end{eqnarray}
where, \mbox{\boldmath$T$} is the transfer matrix and it is defined as,
\begin{eqnarray}
 \mbox{\boldmath$T$}=( \mbox{\boldmath$t$}_0 + \mbox{\boldmath$\tilde{t}$}_0 
 \mbox{\boldmath$t$}_1 +\mbox{\boldmath$\tilde{t}$}_0  
\mbox{\boldmath$\tilde{t}$}_1\mbox{\boldmath$t$}_2+ \hdots +
\mbox{\boldmath$\tilde{t}$}_0  
\mbox{\boldmath$\tilde{t}$}_1 \mbox{\boldmath$t$}_2 \hdots 
\mbox{\boldmath$t$}_n), \nonumber \\ 
\mbox{\boldmath$\tilde{T}$}=( \mbox{\boldmath$\tilde{t}$}_0 + 
\mbox{\boldmath${t}$}_0 
 \mbox{\boldmath$\tilde{t}$}_1 +\mbox{\boldmath$t$}_0  
\mbox{\boldmath$t$}_1\mbox{\boldmath$\tilde{t}$}_2+ \hdots +
\mbox{\boldmath$t$}_0  \mbox{\boldmath$t$}_1 \mbox{\boldmath$\tilde{t}$}_2 
\hdots \mbox{\boldmath$\tilde{t}$}_n).
\end{eqnarray}
After some algebraic calculations we can write from Eq.~\ref{greeneq},
\begin{eqnarray}
\mbox{\boldmath$G$}_{0,0}=[(E+i\eta)  \mbox{\boldmath$I$}-
\mbox{\boldmath$H$}_{0,0}-  \mbox{\boldmath$H$}_{0,1} 
\mbox{\boldmath$T$}]^{-1}.
\end{eqnarray}
With this formalism, the surface Green's function of the left and 
right leads can be found as,
\begin{eqnarray}
\mbox{\boldmath$g$}_{-1,-1}^L&=&[(E+i\eta)  \mbox{\boldmath$I$}-
\mbox{\boldmath$H$}_{0,0}-  \mbox{\boldmath$H$}_{-2,-1}^{\dag} 
\mbox{\boldmath$\tilde{T}$}]^{-1} \nonumber \\
\mbox{\boldmath$g$}_{0,0}^R&=&[(E+i\eta)  \mbox{\boldmath$I$}-
\mbox{\boldmath$H$}_{0,0}-  \mbox{\boldmath$H$}_{0,1} 
\mbox{\boldmath$T$}]^{-1}
\end{eqnarray}
where, $\mbox{\boldmath$H$}_{0,0}$ and $\mbox{\boldmath$H$}_{-2,-1}$ are 
the Hamiltonians for a principal layer in both layer and the tunneling 
matrix between two principal layers in the right-lead, respectively. The 
main advantage of this framework~\cite{meunier,jodar,farajian} rather than 
any other method is that here the number of iterations required for 
convergence is very small~\cite{li,shokri}. Finally, we get the expressions 
for the self-energies of the two leads as,
\begin{eqnarray}
\mbox{\boldmath$\Sigma$}_L&=&\mbox{\boldmath$H$}_{-1,C}^{\dag} 
\mbox{\boldmath$g$}_{-1,-1}^L \mbox{\boldmath$H$}_{-1,C}, \nonumber \\
\mbox{\boldmath$\Sigma$}_R&=&\mbox{\boldmath$H$}_{C,0} 
\mbox{\boldmath$g$}_{0,0}^R \mbox{\boldmath$H$}_{C,0}^{\dag}
\end{eqnarray}
where, $\mbox{\boldmath$H$}_{0,1}$ and $\mbox{\boldmath$H$}_{1,2}$ are 
left lead-to-AGNR and AGNR-to-right lead coupling matrices, respectively. 
Using the above expressions of self-energies for the graphene leads we 
evaluate effective Green's function $\mbox{\boldmath$H$}_{\mbox{\tiny AGNR}}$
with the help of Eq.~\ref{greendevice} and then calculate two-terminal
transmission probability.

\section{Results and Discussion}

In this section we present analytical results of energy band spectrum
for isolated AGNRs and numerical results computed for transmission 
probability through AGNRs under conventional biased conditions. Throughout
our analysis we set the on-site energies in the two side-attached graphene 
leads to zero, $\epsilon_0=0$, and in the AGNR $\epsilon_l=0$ for ordered
sites, while $\epsilon_l=1$eV for defect sites.
The nearest-neighbor coupling strength in the leads ($v_0$) is fixed
at $1$eV, and the coupling parameters $\tau_L$ and $\tau_R$ are also
set at $1$eV. In AGNR, sandwiched between two leads, we use three 
different hopping integrals, $v$, $v^{\prime}$ and $v^{\prime \prime}$, 
and their values are fixed at $1$eV, $0.7$eV and $0.1$eV, respectively.
We fix the equilibrium Fermi energy $E_F$ at zero and choose the units
where $c=e=h=1$. The energy scale is measured in unit of $v$.

\subsection{AGNR without side-attached leads: Energy band structure
and related issues}

To find the energy dispersion relation of an infinitely 
extent (along $x$-direction) AGNR, having a finite width along 
$y$-direction, we establish an effective difference equation analogous 
to the case of an infinite one-dimensional chain. This can be done by proper 
choice of a unit cell (for example, see the unit cell configuration 
presented by the dashed region in Fig.~\ref{model}) from the nano-ribbon. 
With this configuration, the effective difference equation of the AGNR reads,
\begin{equation}
\left(E \mbox{\boldmath$\mathcal{I}$}-\mbox{\boldmath$\mathcal{E}$}\right)
\psi_j=\mbox{\boldmath$\mathcal{T}$}\psi_{j+1} + 
\mbox{\boldmath$\mathcal{T}$}^{\dag} \psi_{j-1} 
\label{diff}
\end{equation}
where,
\begin{eqnarray}
\psi_j= \left(\begin{array}{c}
\psi_{j1} \\
\psi_{j2} \\
\psi_{j3} \\
.\\
.\\
.\\
\psi_{j 2N_y}\end{array}\right).
\end{eqnarray}
In the above relation, {\boldmath${\mathcal{E}}$} and 
{\boldmath${\mathcal{T}}$} correspond to the site-energy 
\begin{figure}[ht] 
{\centering \resizebox*{5.5cm}{8.5cm}{\includegraphics{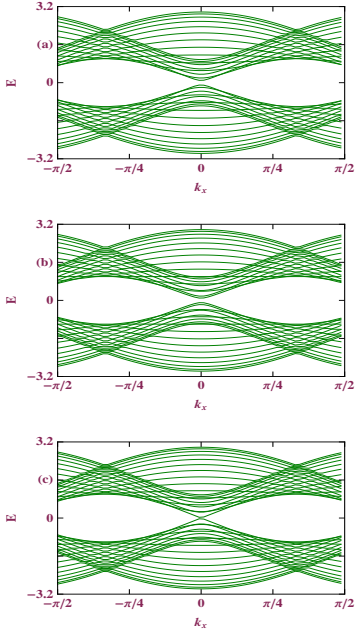}}\par}
\caption{(Color online). Energy band diagrams of armchair graphene 
nanoribbons in absence of any line defect, where (a) $N_y= 18$, 
(b) $N_y=19$, and (c) $N_y=20$.}
\label{ek0}
\end{figure}
\begin{figure}[ht]
{\centering \resizebox*{8.5cm}{3.6cm}{\includegraphics{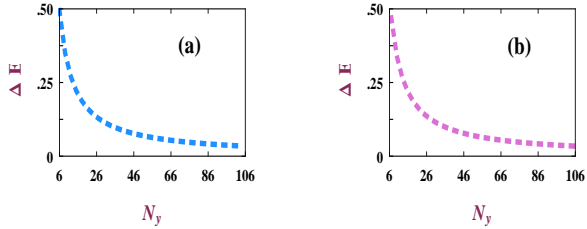}}\par}
\caption{(Color online). Energy gap $\Delta E$, at $k_x=0$, as a function 
of ribbon width $N_y$. (a) $N_y=3n$ and (b) $N_y=3n+1$, where $n$ is an
integer.}
\label{gap_size}
\end{figure}
and nearest-neighbor hopping matrices of the unit cell, respectively, 
and {\boldmath${\mathcal{I}}$} is the identity matrix. The dimension 
of these three matrices is ($2N_y\times2N_y$). Since in the nano-ribbon
translational invariance exists along the $x$-direction, we can write 
$\psi_j$ in terms of the Bloch waves and then Eq.~\ref{diff} gets the form,
\begin{equation}
(E  \mbox{\boldmath$\mathcal{I}$}- \mbox{\boldmath$\mathcal{E}$})=
 \mbox{\boldmath$\mathcal{T}$} e^{ik_x \Lambda}+
 \mbox{\boldmath$\mathcal{T}$}^{\dag} e^{-ik_x \Lambda}
\label{bloch}
\end{equation}
where, $\Lambda=3a$ is the spacing between two neighboring unit cells.
$a$ is the length of each side of hexagonal benzene like ring. Solving
Eq.~\ref{bloch} we find the desired energy dispersion relation ($E$ vs. 
$k_x$) of the armchair ribbon. 

As illustrative examples, in Fig.~\ref{ek0} we present the energy band
diagrams of AGNRs for three different ribbon widths when they are free
from any line defect. In this spectra the three typical numbers ($18$, 
$19$ and $20$) of $N_y$ are chosen only to make $N_y$ in the forms $3n$, 
$3n+1$ and $3n+2$, respectively, since the energy band structures of
\begin{figure}[ht]
{\centering \resizebox*{8.5cm}{7.5cm}{\includegraphics{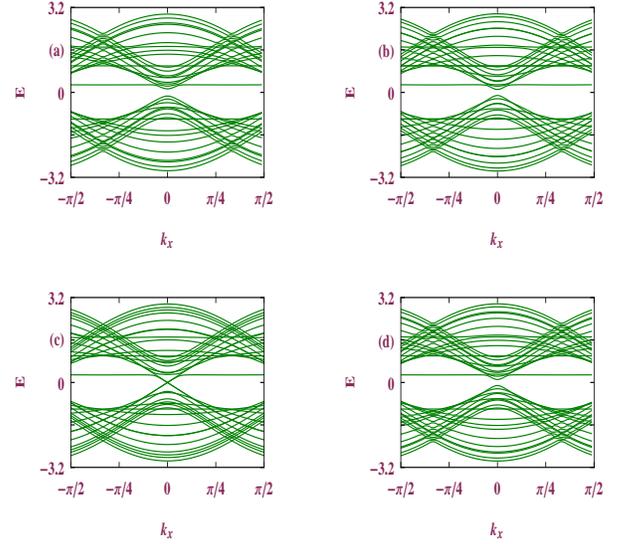}}\par}
\caption{(Color online). Energy band diagrams of armchair graphene 
nanoribbons in presence of line defect. (a) $N_y= 18$, $N_i=5$; 
(b) $N_y=19$, $N_i=4$; (c) $N_y=20$, $N_i=6$ and (d) $N_y=20$, $N_i=5$. 
$N_i$ describes the location of a defect line.}
\label{ek1}
\end{figure}
AGNRs are highly sensitive to these typical ribbon widths~\cite{zheng}.
Here we set $n=6$. From the spectra it is observed that for the particular
case where $N_y=3n+2$, the lowest conduction band and the highest valence 
band coincides with each other at $k_x=0$, resulting zero energy gap in 
the band spectrum (Fig.~\ref{ek0}(c)). This indicates metallic phase
of the AGNR. However, for the other two cases ($N_y=3n$ and $3n+1$), a 
finite gap in the band spectrum is obtained at $k_x=0$ representing the
semiconducting behavior. In these three spectra (Figs.~\ref{ek0}(a)-(c)) 
since $N_y$ is finite, the wavevector along $y$-direction becomes quantized 
and for each value of $k_y$ we get a $E$-$k_x$ curve which results distinct 
energy levels in the $E$-$k_x$ diagram. For large enough $N_y$, energy gaps 
between these energy levels decrease sharply, and therefore, quasi-continuous 
energy bands are formed. The energy gaps between the conduction and valence
bands, at $k_x=0$, of AGNRs with $N_y=3n$ and $N_y=3n+1$ strongly depend
on the value of $n$ i.e., ribbon width. To reveal this fact in 
Fig.~\ref{gap_size} we present the variation of energy gap $\Delta E$ as
a function of $N_y$ for two different cases. It shows that the energy gap
sharply decreases with $N_y$ when $N_y$ becomes smaller, but it eventually
saturates to a finite non-zero value, though it is too small, for large 
enough $N_y$. So, in short, we can emphasize that a metallic phase is 
observed for AGNRs when $N_y=3n+2$, while the semiconducting phase is 
visible for the AGNRs with $N_y=3n$ and $3n+1$.

The results described above for clean nanoribbons i.e., nanoribbons in 
absence of any line defect have already been established in the 
literature, but the central issue of our present investigation - the
interplay between the existence of a line defect, the width of AGNRs
and the location of line defect has not been well addressed earlier. 

To explore it, we present in Fig.~\ref{ek1} the energy band diagrams for
some typical AGNRs in presence of a line defect, where the defect sites
\begin{figure}[ht]
{\centering \resizebox*{5.9cm}{8.6cm}
{\includegraphics{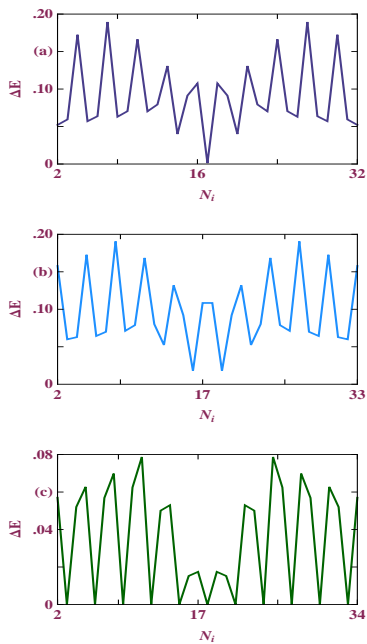}}\par}
\caption{(Color online). Energy gap $\Delta E$, at $k_x=0$, as a 
function of impurity position $N_i$ when (a) $N_y=33$, (b) $N_y=34$, 
and (c) $N_y=35$.}
\label{gap_position}
\end{figure}
are described by the yellow circles as shown in Fig.~\ref{model}. The
location of a line defect is described by the variable $N_i$ and we
assign $N_i=1$ for the edge of an AGNR. In Figs.~\ref{ek1}(a) and (b) the 
results are shown for the ribbon widths $N_y=3n$ and $3n+1$, respectively,
and both for these two cases a finite energy gap around $k_x=0$ is obtained
which reveals the semiconducting nature. The situation is somewhat 
interesting when the width of the AGNR gets the form $3n+2$. The results
are shown in Figs.~\ref{ek1}(c) and (d) where we choose $N_y=20$ 
($=3\times 6 +2$) and locate the defect lines at $6$ and $5$, respectively.
In one case a sharp crossing between the energy levels takes place at 
$k_x=0$, results a metallic phase, while for the other case a finite energy
gap opens up for this typical value of $k_x$ revealing the semiconducting
phase. Thus, a metallic AGNR (width $N_y=3n+2$) can exhibit a metallic
or a semiconducting phase depending on the location of a impurity line in
the ribbon. In a metallic AGNR two types of states, metallic and 
semiconducting, for carbon chains exist which result these two types of
conducting phases depending on the location of line defect, while a
semiconducting AGNR contains only semiconducting chains, and accordingly,
it does not provide the metallic behavior in presence of a line defect. 

The energy gap $\Delta E$, across $k_x=0$, strongly depends on the 
structural details i.e., the location of defect line in the nanoribbon.
\begin{figure}[ht]
{\centering \resizebox*{7cm}{8cm}{\includegraphics{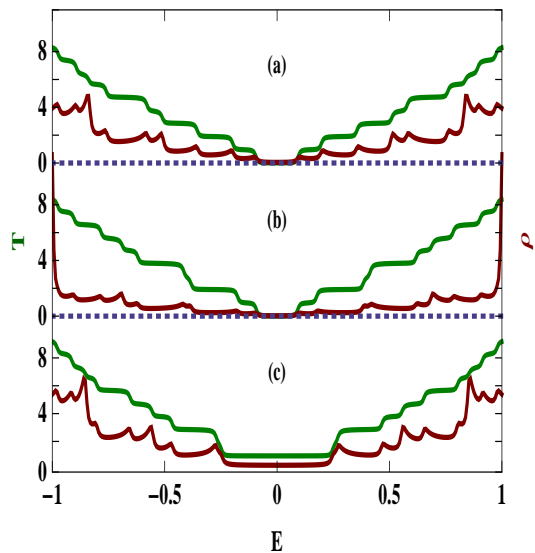}}\par}
\caption{(Color online). Transmission probability (green color) and 
average density of states (red color) for some typical armchair graphene
nanoribbons in absence of line defect, where (a), (b) and (c) correspond
to $N_y=3n$, $3n+1$, and $3n+2$, respectively. Here we set $n=6$ and take
twenty unit cells.}
\label{trans1}
\end{figure}
As illustrative example, in Fig.~\ref{gap_position} we show the variation
of $\Delta E$ as a function of impurity position $N_i$ for three different
ribbon widths. It shows an oscillating behavior with the position of the
defect line. For the semiconducting AGNRs (Figs.~\ref{gap_position}(a) and
(b)) the energy gap never drops to zero (for $N_i=17$ the 
energy gap in Fig.~\ref{gap_position}(a) becomes very small, but still it 
has a finite non-zero value which results a semiconducting phase), while 
for the metallic ribbon (Figs.~\ref{gap_position}(c)) $\Delta E$ exactly
vanishes when $N_i$ becomes equal to $3p$, $p$ being an integer. These 
phenomena promote a design concept based on the structural details as 
semiconducting devices with variable energy band gaps.

\subsection{AGNR with side-attached leads: Two-terminal transmission 
probability and ADOS}

Keeping in mind a possible experimental realization of the system, we
clamp a finite sized armchair graphene nanoribbon between two ideal 
semi-infinite graphene leads (the left and right leads) making a 
lead-AGNR-lead bridge (see Fig.~\ref{model}). Below we present our
numerical results for average density of states (ADOS) and two-terminal
transmission probability through finite sized AGNRs under conventional
biased conditions.

In Fig.~\ref{trans1} we show the variation of transmission probability, 
$T$, together with the average density of states (ADOS), $\rho$, as a 
\begin{figure}[ht]
{\centering \resizebox*{8.5cm}{6.5cm}{\includegraphics{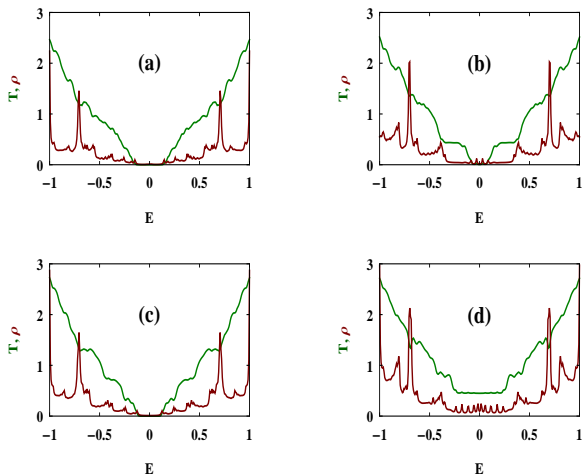}}\par}
\caption{(Color online). Transmission probability (green color) and
average density of states (red color) for some typical armchair graphene
nanoribbons in presence of single line defect. In (a) and (b) the results
are shown for $N_y=3n$ and $3n+1$, respectively, considering $N_i=5$,
while in (c) and (d) the results are given for $N_y=3n+2$ when the defect
line is placed at $5$-th ($N_i=5$) and $6$-rd ($N_i=6$) lines, respectively.
Here we take twenty unit cells and fix $n$ at $6$.}
\label{trans2}
\end{figure}
function of energy $E$ for some typical finite sized AGNRs in the absence 
of any line defect, where (a), (b) and (c) correspond to the results for 
$N_y=3n$, $3n+1$ and $3n+2$, respectively. Here we set $n=6$ and choose
twenty unit cells. Our previous analytical arguments for the AGNRs are
exactly corroborated in these diagrams. A finite energy gap in the
transmission probability associated with the energy gap in ADOS spectrum
is obtained when $N_y$ becomes identical to $3n$ and $3n+1$, as shown in
Figs.~\ref{trans1}(a) and (b). This behavior emphasizes the semiconducting 
phase for these ribbon widths. On the other hand a gap less spectrum is 
observed when $N_y$ becomes equal to $3n+2$ (Fig.~\ref{trans1}(c)), which
indicates the metallic phase of the AGNR.

Finally, we describe the results shown in Fig.~\ref{trans2}, where we 
present transmission probability and ADOS for some typical AGNRs in 
presence of single line defect. In (a) and (b) the results are presented 
for $N_y=3n$ and $3n+1$, respectively, and for both these two cases 
transmission function shows a finite energy gap across $E=0$ exhibiting
the semiconducting nature, with reduced amplitude compared to the spectra 
given in Fig.~\ref{trans1}. The structural dependence on the conducting 
behavior in metallic AGNR ($N_y=3n+2$) in presence of line defect is 
clearly visible from the spectra given in Figs.~\ref{trans2}(c) and (d),
where the defect lines are placed in the $4$-th and $3$-th lines, 
respectively. For the first case, it provides the semiconducting behavior,
while in the other case the metallic phase is obtained, which perfectly
corroborate our previous analytical findings.

\section{Conclusion}

To summarize, we have investigated in detail the characteristics of 
energy band spectrum of armchair graphene nanoribbons in presence of
line defect within a simple non-interacting tight-binding framework.
The essential results have been presented in two parts. In the first
part we have presented analytical results of energy band spectrum for
isolated AGNRs. From our analytical results we have analyzed that 
depending on the location of a line defect a metallic AGNR can provide
either a metallic or a semiconducting phase, while a semiconducting
AGNR provides only the semiconducting phase with variable band gap.
In the second part, we have discussed numerical results for transmission
probability together with ADOS, keeping in mind a possible experimental 
realization of the system. We have shown that our numerical results
exactly corroborate the analytical findings. Though the results presented 
in this article are worked out at absolute zero temperature limit, the 
results should remain valid even at finite temperatures ($\sim300\,$K)
since the broadening of the energy levels of the AGNR due to its coupling 
with the metal leads is much higher than that of the thermal 
broadening~\cite{datta1,we2,we3,we4,we5,we6,we7,we8}.

Throughout our work, we have addressed the electronic
transport properties in AGNRs for some typical parameter values. In 
our model calculations we chose them only for the sake of simplicity. 
Though the results presented here change numerically with these parameter 
values, but all the basic features remain exactly invariant.

\end{document}